\documentclass [12pt]{article}
\def\be{\begin{equation}}
\def\eea{\end{eqnarray}}
\def\bea{\begin{eqnarray}}
\def\ee{\end{equation}}

\usepackage[psamsfonts]{amssymb}
\usepackage{amsmath}
\author{F. Kheirandish$^{1}$ \footnote{fardin$_{-}$kh@phys.ui.ac.ir}
\\ $^{1}$ {\small Department of Physics, University of Isfahan,}
\\ {\small Hezar Jarib Ave., Isfahan, Iran.}}
\title{Husimi distribution function and one-dimensional Ising model}
\date{}
\begin{document}
\maketitle
\begin{abstract}
\noindent Husimi distribution function for the one-dimensional
Ising model is obtained. One-point and joint distribution
functions are calculated and their thermal behaviour are
discussed.
\end{abstract}
\section{Introduction}
Husimi distribution function \cite{husimi}, is a positive definite
distribution over the parameter space of a definite set of
coherent states \cite{zhang}. Positive definitness of the Husimi
distribution function, makes it as a good candidate for
investigating quantum features using a classical like
distribution over the phase space. In this case, to each coherent
state $|z\rangle$ of the system there corresponds a unique
distribution function $\mu(z)=\frac{1}{Z}\langle
z|e^{-\beta\hat{H}}|z\rangle$, where $Z$, is the partition
function of the system. $\hat{H}$ and $\beta$ are Hamiltonian and
inverse temperature parameter respectively. In this paper, using
the single-fermion or spin-$\frac{1}{2}$ coherent states, the
coherent states of a one-dimensional spin chain are obtained. The
Husimi distribution function is then obtained for the
one-dimensional Ising model as the prototype of some important
models in statistical mechanics. One-point and joint distribution
functions or correlation functions are calculated and their
thermal behaviour is discussed.
\section{One-dimensional Ising model}
Hamiltonian of the one dimensional Ising model in a homogeneous
magnetic field $B$ along the $z$-axis is
\begin{equation}\label{f 20}
\hat{H} =
-J\sum_{k=1}^{N}\hat{S}_{k,z}\hat{S}_{k+1,z}+B\sum_{k=1}^{N}\hat{S}_{k,z},
\end{equation}
where periodic boundary condition is assumed, i.e, $N+1\equiv 1.$
The Hilbert space $\aleph$ of this model is a $2^{N}$-dimensional
vector space spaned by the following tensor products as basis
vectors \be
|\vec{i}\rangle=|i_{1}\rangle\otimes|i_{2}\rangle\otimes...\otimes|i_{N}\rangle=
\prod_{k=1}^{N}\otimes|i_{k}\rangle, \ee where
$|i_{k}\rangle\in\{|+\rangle, |-\rangle\}$, for $k=1...N$. The
basis vectors $|\vec{i}\rangle$, are Hamiltonian eigenvectors
with respective eigenvalues $E_{\vec{i}}$, \bea
\hat{H}|\vec{i}\rangle&=&(-J\sum_{k=1}^{N}i_{k}i_{k+1}+B\sum_{k=1}^{N}i_{k})|\vec{i}\rangle,\nonumber\\
E_{\vec{i}}&=&-J\sum_{k=1}^{N}i_{k}i_{k+1}+B\sum_{k=1}^{N}i_{k},
\eea
\section{Ising model coherent states}
Coherent states of the one dimensional Ising model can be defined
from the tensor product of the single-fermion coherent states
[2]. A single-fermion coherent state for the spin half
$(s=\frac{1}{2})$ representation can be written like this \be
|z\rangle=\sin(\frac{\theta}{2})\exp(-i\varphi)|+\rangle+\cos(\frac{\theta}{2})|-\rangle,
\ee where $z=(\frac{\theta}{2})exp(-i\varphi)$,
$0\leq\theta\leq\pi$, $0\leq\varphi\leq 2\pi$ and $|+\rangle$,
$|-\rangle$ are $\hat{S}_z$ eigenvectors. The completeness of the
fermion coherent states is \be \frac{1}{2\pi}\int
d\Omega|z\rangle\langle z|=1, \ee where $d\Omega=\sin\theta
d\theta d\varphi$. Tensor product of these fermion coherent states
makes Ising model coherent states \be
|\vec{z}\rangle=\prod_{k=1}^{N}\otimes(\sin(\frac{\theta_{k}}{2})\exp(-i\varphi_{k})|+\rangle
+\cos(\frac{\theta_{k}}{2})|-\rangle), \ee the scalar product
between a coherent state $|\vec{z}>$ and an eigenvector
$|\vec{i}>$, of the Hamiltonian $\hat{H}$, is  \be
\langle\vec{z}|\vec{i}\rangle=
\prod_{k=1}^{N}(\sin(\frac{\theta_{k}}{2})\exp(-i\varphi_{k})\delta_{i_{k},1}+
\cos(\frac{\theta_{k}}{2})\delta_{i_{k},-1}), \ee so \be
|\langle\vec{z}|\vec{i}\rangle|^{2}=
\prod_{k=1}^{N}(\sin^{2}(\frac{\theta_{k}}{2})\delta_{i_{k},1}+
\cos^{2}(\frac{\theta_{k}}{2})\delta_{i_{k},-1}).\ee Partition
function of the Hamiltonian (1) is defined as \be
Z=\sum_{\vec{i}}\langle\vec{i}\mid \exp(-\beta
\hat{H})\mid\vec{i}\rangle=tr(\hat{T}^{N}), \ee
 where \be
\hat{T}=\begin{pmatrix}
  \exp(\beta(J-B)) & \exp(-\beta J) \\
  \exp(-\beta J) & \exp(\beta (J+B))
\end{pmatrix},
\ee is the transition matrix.
\section{Husimi distribution function}
Husimi distribution function is a normalized positive definite
distribution in the parameter space of coherent states and is
defined as follows \bea
\mu(\vec{z})&=&\frac{1}{Z}\langle \vec{z}|\exp(-\beta\hat{H})|\vec{z}>,\nonumber\\
&=&\frac{1}{Z}\sum_{\{\vec{i}\}}\langle\vec{z}|\exp(-\beta\hat{H})|\vec{i}\rangle\langle
\vec{i}|\vec{z}>,\nonumber\\
&=&\frac{1}{Z}\sum_{\{\vec{i}\}}e^{\beta
J\sum_{k=1}^{N}i_ki_{k+1}-\beta
B\sum_{k=1}^{N}i_k}|\langle\vec{z}|\vec{i}\rangle|^2, \eea
substituting (8) in (11) and doing some routine calculations, we
get the following relation for Husimi distribution \bea \mu
(\vec{z})&=&\frac{1}{2^N}\{1-\sum_{m=1}^{N}\langle
\hat{S}_{m,z}\rangle u_m+\sum_{m<n}
\langle \hat{S}_{m,z}\hat{S}_{n,z}\rangle u_m u_n+\cdots\nonumber\\
&+&(-1)^{N} \langle \hat{S}_{1,z}\hat{S}_{2,z}\cdots
\hat{S}_{N,z}\rangle u_1u_2\cdots u_n\}, \eea where for simplicity
$u_k=cos\theta_k$ is assumed and a general $n$-point corelation
function $\langle \hat{S}_{m,z} \hat{S}_{n,z}\cdots
\hat{S}_{r,z}\rangle$ is defined as \be \langle \hat{S}_{m,z}
\hat{S}_{n,z}\cdots
\hat{S}_{r,z}\rangle=\frac{1}{Z}\sum_{\vec{i}}i_m i_n\cdots i_r
e^{\beta J\sum_{k=1}^{N}i_k i_{k+1}-\beta B\sum_{k=1}^{N}i_k},
\ee where $\hat{S}_{k,z}|i_k\rangle=i_k|i_k\rangle$, from (12) it
is clear that the Husimi distribution function is independent of
parameters $\varphi_k$ and contains the information about all the
$n$-point correlation functions, i.e., it is the generator of the
$n$-point functions of the Ising model \bea \langle
\hat{S}_{r_1,z}\hat{S}_{r_2,z}\cdots
\hat{S}_{r_m,z}\rangle&=&(-3)^{m}\int\cdots\int\mu(\vec{z})u_{r_1}u_{r_2}\cdots
u_{r_m}\prod_{k=1}^{N}du_{k},\nonumber\\
&=&(-3)^{m}\langle u_{r_1}u_{r_2}\cdots u_{r_m}\rangle. \eea Now
the Husimi distribution function related to the cite $k$, can be
calculated by integrating over parameters of the phase spase
except those belong to the $k$th cite, so \be
\mu(u_k)=\frac{1}{2}(1-u_k\frac{tr[\hat{T}^{N}\sigma_{z}]}{tr[\hat{T}^{N}]}),
\ee where as usual $\sigma_{z}$ is \be \sigma_{z}=\begin{pmatrix}
  1 & 0 \\
  0 & -1
\end{pmatrix}.
\ee The joint Husimi distribution function related to cites $m$
and $n$, is equal to \be \mu(u_m,
u_n)=\frac{1}{4}(1-(u_m+u_n)\frac{tr[\hat{T}^{N}\sigma_{z}]}{tr[\hat{T}^{N}]}+u_m
u_n\frac{tr[\hat{T}^{N-n+m}\sigma_z\hat{T}^{n-m}\sigma_z]}{tr[\hat{T}^{N}]}),
\ee for calculating $tr[\hat{T}\sigma_z]$ and
$tr[\hat{T}^{N-n+m}\sigma_z\hat{T}^{n-m}\sigma_z]$ we can
diagonalize the matrix $\hat{T}$, and write the matrix $\sigma_z$
in the basis of eigenvectors of $\hat{T}$, the eigenvalues of
$\hat{T}$ are \be \lambda_{\pm}=\exp(\beta J)\cosh(\beta
B)\pm\sqrt{\exp(2\beta J)\cosh^{2}(\beta B)-2\sinh(\beta J)}, \ee
and the transformation matrix between the standard basis
$\{|+\rangle, |-\rangle\}$ and the eigenvectors of $\hat{T}$ is
\be U=\begin{pmatrix}
  \sin(\omega) & -\cos(\omega) \\
  \cos(\omega) & \sin(\omega)
\end{pmatrix},
\ee where $|\lambda_{+}\rangle=\begin{pmatrix}
  sin(\omega) \\
  cos(\omega)
\end{pmatrix}$ and $|\lambda_{-}\rangle=\begin{pmatrix}
  -cos(\omega) \\
  sin(\omega)
\end{pmatrix}$ are normalized eigenvectors of $\hat{T}$ belonging to eigenvalues
$\lambda_{+}$ and $\lambda_{-}$ respectively and \be
\tan(\omega)=\frac{\exp(-\beta J)}{\exp(\beta J)\sinh(\beta
B)+\sqrt{\exp(2\beta J)\sinh^{2}(\beta B)+\exp(-2\beta J)}}. \ee
Now in the basis $|\lambda_{+}\rangle$, $|\lambda_{-}\rangle$, we
have \bea tr[\hat{T}\sigma_{z}]&=&tr(\begin{pmatrix}
  \lambda_{+}^N & 0 \nonumber\\
  0 & \lambda_{-}^N
\end{pmatrix}\begin{pmatrix}
  -\cos(2\omega) & -\sin(2\omega) \nonumber\\
  -\sin(2\omega) & \cos(2\omega)
\end{pmatrix}),\\
&=&tr[\begin{pmatrix}
  -\lambda_{+}^{N}\cos(2\omega) & -\lambda_{+}^{N}\sin(2\omega)\nonumber\\
  -\lambda_{-}^{N}\sin(2\omega) & \lambda_{-}^{N}\cos(2\omega)
\end{pmatrix}],\nonumber\\
&=&(\lambda_{-}^{N}-\lambda_{+}^{N})\cos(2\omega), \eea and
similarly \bea
tr[\hat{T}^{N-j+i}\sigma_{z}\hat{T}^{j-i}\sigma_{z}]&=&
\lambda_{+}^{N}\cos^{2}(2\omega)+\lambda_{+}^{N-j+i}\lambda_{-}^{j-i}\sin^{2}(2\omega)\nonumber\\
&+&\lambda_{-}^{N-j+i}\lambda_{+}^{j-i}\sin^{2}(2\omega)+\lambda_{-}^{N}\cos^{2}(2\omega),
\eea substituting (21) and (22) in (15) and (17) respectively, we
obtain \be
\mu(u_{k})=\frac{1}{2}(1-\cos(2\omega)[\frac{\lambda_{-}^N-\lambda_{+}^N}
{\lambda_{-}^N+\lambda_{+}^N}]u_k), \ee \bea
\mu(u_i,u_j)&=&\frac{1}{4}(1-(u_i+u_j)\cos(2\omega)[\frac{\lambda_{-}^N-\lambda_{+}^N}
{\lambda_{-}^N+\lambda_{+}^N}]u_k
+u_iu_j\cos^{2}(2\omega)\nonumber\\
&+&u_iu_j\sin^{2}(2\omega)\frac{\lambda_{+}^{N-j+i}\lambda_{-}^{j-i}+
\lambda_{-}^{N-j+i}\lambda_{+}^{j-i}}{\lambda_{-}^N+\lambda_{+}^N}),
\eea in thermodynamic limit, i.e., $N\rightarrow\infty$, we
have \bea \mu(u_k)&=&\frac{1}{2}(1+\cos(2\omega)u_k),\nonumber\\
\mu(u_i,u_j)&=&\frac{1}{4}(1+(u_i+u_j)\cos(2\omega)+
u_iu_j\cos^{2}(2\omega)\nonumber\\
&+&u_iu_j(\frac{\lambda_{-}}{\lambda_{+}})^{j-i}\sin^{2}(2\omega)),
\eea such that \bea \cos(2\omega)&=&\frac{2\exp(\beta
J)\sinh(\beta B)}{\sqrt{4\exp(2\beta J)\sinh^{2}(\beta
B)+4\exp(-2\beta
J)}},\nonumber\\
&=&\frac{Sgn(B)}{\sqrt{1+\frac{\exp(-4\beta J)}{\sinh^{2}(\beta
B)}}},\nonumber\\ \tan(2\omega)&=&\frac{exp(-2\beta
J)}{\sinh(\beta B)}, \eea where $Sgn(B)$, is the sign function.
 The behaviour of one-point distribution function $\mu(u_k)$, in
 high and low temperatures, is
  \be
  \mu(u_k)=
  \begin{cases}
    \frac{1}{2} & \text{$\beta\rightarrow 0$}, \\
    \frac{1}{2}(1+u_k) & \text{$\beta\rightarrow\infty$},
  \end{cases}
\ee
  in zero temperature, $\mu(u_k)$ attains it's maximum for
  $\theta_k=0$, which from (4) corresponds to the spin down state,
  $|-\rangle$, as expected.\\
  Similarly we can find the following
  behaviour for joint distribution $\mu(u_i,u_j)$,
  \be
  \mu(u_i,u_j)=
  \begin{cases}
    \frac{1}{4} & \text{$\beta\rightarrow 0$}, \\
    \frac{1}{4}(1+u_i+u_j+u_iu_j) &
    \text{$\beta\rightarrow\infty$},
  \end{cases}
  \ee
in this case the disjoint distribution attains it's maximum value
for $\theta_i=\theta_j=0$, which means that most probably the two
spins are in the spin down state, i.e., $|-\rangle$, in derivation
of these behaviours it is assumed that $Sign(B)=1$, by reversing
the magnetic field $(Sign(B)=-1)$, the spins flip to the spin up
state.

\end{document}